\documentclass[11pt]{article}
\usepackage{amsmath,amssymb,lmodern}
\usepackage[T1]{fontenc}
\let\el\l
\let\ot\otimes
\let\ov\overline
\let\os\overrightarrow
\let\pa\partial
\let\q\quad
\def\qh#1{\quad\hbox{#1}\quad}
\let\ul\underline
\let\wh\widehat
\let\wt\widetilde
\let\a\alpha
\let\b\beta
\let\d\delta
\let\ve\varepsilon
\let\g\gamma
\let\G\varGamma
\let\l\lambda
\let\La\Lambda
\let\o\omega
\let\O\varOmega
\let\vf\varphi
\let\si\sigma
\let\Si\varSigma
\let\t\theta
\let\T\varTheta
\let\z\zeta
\def\m{{\mu\nu}}
\def\cD{\mathcal D}
\def\cL{\mathcal L}
\def\C{\mathbb C}
\def\dg #1,#2,#3,{{{#1}_{#2}}^{\!#3}}
\def\gd #1,#2,#3,{{#1^{#2}}_{\!#3}}
\def\nad#1#2{\overset{\rm #1}{#2}}
\def\hor{\operatorname{hor}}
\def\ver{\operatorname{ver}}
\def\id{\operatorname{id}}
\def\ad#1{\operatorname{ad}_{#1}}
\def\Spin{{\rm Spin}}
\def\SO{{\rm SO}}

\let\ti\textit
\def\beq{\begin{equation}\label}
\def\bea{\begin{eqnarray}\label}
\def\bml{\begin{multline}\label}
\def\bg{\begin{gather}\label}
\let\bal\aligned \let\eal\endaligned
\let\bga\gathered \let\ega\endgathered
\let\la\lambda
\def\U{\text{U}}
\let\Ps\varPsi
\let\PS\varPsi
\def\(#1){{(#1)}}
\def\gdv{\nad{gauge}d}
\def\pl{_{\rm pl}}

\def\ct{constant}
\def\cd{coordinate}
\def\di{dimension}
\def\spt{space-time}
\def\pt{potential}
\def\cf{coefficient}
\def\cn{connection}
\def\dv{derivative}
\def\elm{electromagneti}
\def\e{equation}
\def\f{function}
\def\gr{gravitation}
\def\nos{nonsymmetric}
\def\NK#1{Nonsymmetric Kaluza--Klein #1Theory}
\def\rf{reflection}
\def\so{solution}
\def\st{such that }
\def\s{symmetric}
\def\tf{transformation}
\def\wrt{with respect to }
\def\Pl{Planck}

\author{M. W. Kalinowski\\
Pracownia Bioinformatyki, Instytut Medycyny
Do\'swiadczalnej i~Klinicznej PAN,\\
ul. Pawi\'nskiego 5, 02-106 Warszawa, Poland\\
e-mail:  markwkal1@gmail.com}
\title{Dirac equation\\
in the Nonsymmetric Kaluza--Klein Theory}
\date{}
\begin{document}
\maketitle
\begin{abstract}
We redrive Dirac equation in the Nonsymmetric Kaluza--Klein Theory getting
an electric dipole moment of fermion and CP violation.
\end{abstract}

In the paper we deal with a generalization of a Dirac \e\ on~$P$ (a~metrized
\elm c fiber bundle, see Refs \cite1--\cite5). Some elements of geometry are
given in Appendices A and~B. Thus we
consider spinor fields $\Ps,\ov\Ps$ on~$P$ transforming according to
$\Spin(1,4)$ (a~double covering group of $\SO(1,4)$---de Sitter group). We
want to couple these fields to gravity and electromagnetism. For
$\Ps$ and $\ov\Ps$ we have $\Ps,\ov\Ps:P\to\C^4$ and
\beq{4.1}
\bal
\Ps(\vf(g)p)&=\si(g^{-1})\Ps(p)\\
\ov\Ps(\vf(g)p)&=\ov\Ps(p)\si(g),
\eal
\end{equation}
where $\si\in\cL(\C^4)$, $p=(x,g_1)\in P$, $g,g_1\in\U(1)$.

On $E$ we define spinor ordinary fields $\psi,\ov\psi:E\to\C^4$. We suppose
that $\psi$ and~$\ov\psi$ are defined up to a phase factor and that
\beq{4.2}
\bal
\psi^f(x)&=\Ps(f(x))\\
\ov\psi{}^f(x)&=\ov\Ps(f(x))
\eal
\end{equation}
where $f:E\to P$ is a section of a bundle $\ul P$. In some sense spinor fields
on~$P$ are lifts of spinors on~$E$ (see Appendix B),
\beq{4.3}
\bal
\Ps(f(x))=\pi^\ast(\psi^f(x)),\quad&\psi^f=f^\ast \Ps\\
\ov\Ps(f(x))=\pi^\ast(\ov\psi{}^f(x)),\quad&\ov\psi{}^f=f^\ast \ov\Ps.
\eal
\end{equation}

Let us consider a different section of a bundle $\ul P$, $e:E\to P$. In this
case we have
$$
\gathered
\psi^e=e^\ast \Ps, \q \ov \psi=e^\ast\ov\Ps, \q
\psi{}^e(x)=\Ps(e(x)), \q \ov\psi{}^e(x)=\ov\Ps(e(x)),\\
\psi^e(x)=\psi^f(x)\exp\Bigl(\frac{ikq}{\hbar c}\,\chi(x)\Bigr), \q
\ov\psi{}^e(x)=\ov\psi{}^f(x)\exp\Bigl(-\frac{ikq}{\hbar c}\,\chi(x)\Bigr),
\endgathered
$$
where $kq$ is a charge of a fermion, $k=0,\pm1,\pm2, \dots$, for an electron $k=1$, $\chi$ is a gauge
changing \f.

Let us define an exterior gauge \dv\ $\gdv$ of the field~$\Ps$. One gets
\beq{4.4}
d\Ps=\z_\mu \Ps\t^\mu+\z_5\Ps\t^5
\end{equation}
and
\beq{4.5}
\bal
\gdv\Ps&=\hor d\Ps=\z_\mu\Ps\t^\mu\\
\gdv\ov\Ps&=\hor d\ov\Ps=\z_\mu\ov\Ps\t^\mu.
\eal
\end{equation}
Let $\g_\mu\in\cL(\C^4)$ be Dirac's matrices obeying the conventional
relations
\beq{4.6}
\{\g_\mu,\g_\nu\}=2\eta_\m
\end{equation}
(where $\eta_\m$ is a Minkowski tensor of signature $(---+)$) and let $B=B^+$
be a matrix \st
\beq{4.7}
\g^{\mu+}=B\g^\mu B^{-1}, \quad \ov\psi=\psi^+ B
\end{equation}
(the indices are raised by $\eta^\m$, an inverse tensor of $\eta_\m$), where
``$+$'' is a Hermitian conjugation, and
\beq{4.8}
\si_\m=\tfrac18[\g_\mu,\g_\nu].
\end{equation}
We define
$$
\g^5=\g^1\g^2\g^3\g^4 \in \cL(\C^4).
$$
One can easily check that
\beq{4.9}
\{\g_A,\g_B\}=2\ov g_{AB}
\end{equation}
where
\beq{4.10}
\begin{array}{rcl}
{}&\ov g_{AB}={\rm diag}(-1,-1,-1,+1,-1)&\\[2pt]
\hbox{and }&\g^A=(\g^\a,\g^5)&
\end{array}
\end{equation}
(the indices are raised by $\ov g{}^{AB}$, an inverse tensor of $\ov g_{AB}$).
We have
\beq{4.11}
\g^{5+}=B\g^5B^{-1} \qh{and} \ov\Ps=\PS^+B.
\end{equation}
So
\beq{4.12}
\g^{A+}=B\g^A B^{-1}.
\end{equation}

On the manifold $P$ we have an orthonormal \cd\ system $\t^A$ and we can
perform an infinitesimal change of the frame
\beq{4.13}
\bga
\t^{A'}=\t^A+\d \t^A=\t^A-\gd \ve,A,B,\t^B\\
\ve_{AB}+\ve_{BA}=0.
\ega
\end{equation}
If the spinor field $\Ps$ corresponds to $\t^A$ and $\Ps'$ to $\t^{A'}$ then
we get
\beq{4.14}
\bal
\Ps'&=\Ps+\d\Ps=\Ps-\ve^{AB}\wh\si_{AB}\Ps\\
\ov\Ps'&=\ov\Ps+\d\ov\Ps=\ov\Ps+\ov\Ps\wh\si_{AB}\ve^{AB}
\eal
\end{equation}
($\Ps$ and $\ov\Ps$ are Schouten $\si$-quantities (see Refs
\cite{433},~\cite{434}) where
\beq{4.15}
\wh\si_{AB}=\tfrac18[\g_A,\g_B].
\end{equation}
Notice that the \di\ of the spinor space for a $2n$-\di al space is $2^n$ and
it is the same for a $(2n+1)$-\di al one (in our case $n=2$).

We take a spinor field for a 5-dimensional space $P$ and assume that the
dependence on the 5th \di\ is trivial, i.e.\ Eq.~\eqref{4.1} holds. Taking a
section we obtain spinor fields on~$E$.

Let us introduce some new notions. We introduce a Levi-Civita symbol and a
dual Cartan's base
\bea{4.16}
{}&\ov\eta_{\a\b\g\d}, \qquad \ov\eta_{1234}=\sqrt{-\det(g_\(\a\b))}\\
&\ov\eta_\a=\frac1{2\cdot3}\,\ov\t^\d\wedge\ov\t^\g\wedge\ov\t^\b
\ov\eta_{\a\b\g\d} \label{4.17}\\
&\ov\eta=\frac14\ov\t^\a\wedge\ov\eta_\a.
\end{eqnarray}
We define
\beq{4.19}
\bal
\eta_\a&=\pi^\ast (\ov\eta_\a)\\
\eta&=\pi^\ast(\ov\eta)
\eal
\end{equation}
We rewrite here a Riemannian part of the \cn\
\[
\gd w,A,B,=\left( \begin{array}{c|c}
\pi^*(\gd \ov w,\a,\b,)+g^{\g\a}H_{\g\b}\t^5\  & \ H_{\b\g}\t^\g
\vrule height0pt depth 5pt width0pt \\
\hline
g^{\a\b}(H_{\g\b}+2F_{\b\g})\t^\g\ &\ 0
\vrule height10pt depth 0pt width0pt
\end{array} \right)
\]
where $H_{\b\g}$ is a tensor on $E$ \st
\[
g_{\d\b}g^{\g\d}H_{\g\a}+g_{\a\d}g^{\d\g}H_{\b\g}=
2g_{\a\d}g^{\d\g}F_{\b\g},
\]
introducing the \ct\
$\l=\frac{2\sqrt{G_N}}{c^2}$,
\beq{4.20}
\gd\wt w{},A,B,=\left(
\begin{array}{c|c}
\pi^\ast\bigl(\gd{\wt{\ov w}}{},\a,\b,)+\frac\l2\pi^\ast(\gd F,\a,\b,)\t^5\, &\,
\frac\l2 \pi^\ast(\gd F,\a,\g,\ov \t{}^\g)\\[2pt]\hline
\noalign{\vskip2pt}
-\frac\l2\pi^\ast (F_{\b\g}\ov \t{}^\g) & 0
\end{array}
\right)
\end{equation}
(see Refs \cite{98}, \cite{420}).

Let us consider exterior covariant \dv s of spinors $\Ps$ and $\ov\Ps$,
\beq{4.21}
\bal
\wt D\Ps&=d\Ps+\gd\wt w{},A,B,\dg\wh\si{},A,B,\Ps\\
\wt D\ov\Ps&=d\ov\Ps-\gd\wt w{},A,B,\Ps \dg\wh\si{},A,B,
\eal
\end{equation}
\wrt the Riemannian \cn\ $\gd \wt w{},A,B,$.

Now we introduce a \dv\ $\cD$, i.e.\ an exterior ``gauge'' \dv\ of a new
kind. This \dv\ may be treated as a generalization of minimal coupling scheme
between spinor and \elm c field on~$P$,
\beq{4.22}
\bal
\cD\Ps&=\hor D\Ps\\
\cD\ov\Ps&=\hor D\ov\Ps.
\eal
\end{equation}
We get
\beq{4.23}
\bal
\cD\Ps&=\wt{\ov\cD}\Ps-\frac\l8\,\gd F,\a,\mu,[\g_\a,\g_5]\Ps\t^\mu\\
\cD\ov\Ps&=\wt{\ov\cD}\,\ov\Ps+\frac\l8\,\gd F,\a,\mu,\ov\Ps[\g_\a,\g_5]\t^\mu
\eal
\end{equation}
where
\beq{4.24}
\bal
\wt{\ov\cD}\Ps&=\gdv \Ps+\pi^\ast (\gd{\wt{\ov w}}{},\a,\b,)\dg\si,\a,\b,\Ps\\
\wt{\ov\cD}\ov\Ps&=\gdv \ov\Ps-\pi^\ast (\gd{\wt{\ov w}}{},\a,\b,)
\ov\Ps\dg\si,\a,\b,.
\eal
\end{equation}

The \dv\ $\wt{\ov\cD}$ is a covariant \dv\ \wrt both $\pi^\ast(\gd{\wt{\ov
w}}{},\a,\b,)$ and ``gauge'' at once. It introduces an interaction between
\elm c and \gr al fields with Dirac's spinor in a classical well-known way
($\wt{\ov\cD}\Ps =\hor\wt{\ov D}\Ps$).

In Dirac theory we have the following Lagrangian for a spinor $\frac12$-spin
field on~$E$:
\beq{4.25}
\cL(\psi,\ov\psi,d)=i\,\frac{\hbar c}2\bigl(\ov\psi\, \ov l\wedge d\psi
+d\ov\psi\wedge \ov l\psi\bigr)+mc^2\ov\psi\psi\ov\eta
\end{equation}
where $\ov l=\g_\mu\ov\eta{}^\mu$.

Let us lift Lagrangian on a manifold $P$. We pass from spinors $\psi$ and
$\ov \psi$ to $\Ps$ and $\ov\PS$ and from the \dv\ $d$ to $\gdv$ or
to~$\wt{\ov \cD}$. This is a classical way. Moreover, we have to do with a
theory which unifies gravity and \elm sm and in order to get new physical
effects we should pass to our new \dv~$\cD$. Simultaneously we pass from
$\ov\eta$ to~$\eta$ and from $\ov l$ to $\pi^\ast(\ov l)=l$.

In this way one gets
\beq{4.26}
\cL_D(\Ps,\ov\Ps,\cD)=\frac{i\hbar c}2\bigl(\ov\PS l\wedge\cD\Ps
+\cD\ov\Ps\wedge l\Ps\bigr)+mc^2\ov\Ps \Ps\eta.
\end{equation}
Using formulae \eqref{4.23} one obtains
\beq{4.27}
\cL_D(\Ps,\ov\Ps,\cD)=\cL_D(\Ps,\ov\Ps,\wt{\ov \cD})
-i\,\frac{2\sqrt {G_N}}c\,\hbar F_\m \ov\Ps\g_5\si^\m \Ps\eta
\end{equation}
where
\beq{4.28}
\cL_D(\Ps,\ov\Ps,\wt{\ov \cD})=\frac{i\hbar c}2\bigl(\ov\Ps l\wedge
\wt{\ov\cD} \Ps+\wt{\ov\cD}\ov\Ps\wedge l\Ps\bigr)+mc^2\ov\Ps \Ps\eta.
\end{equation}

Now we should go back to a \spt\ $E$ (see Appendix~B) and we get the
following Lagrangian
\bg{4.29a}
\cL_D(\psi,\ov\psi,\cD)=\cL_D(\psi,\ov\psi,\wt{\ov\cD})
-i\,\frac{2\sqrt{G_N}}c\,\hbar F^\m\ov\psi\g_5\si_\m \psi\\
\cL_D(\psi,\ov\psi,\wt{\ov\cD})=\frac{i\hbar c}2 \bigl(\ov\psi\,\ov l
\wedge \wt{\ov\cD}\psi+\wt{\ov\cD}\psi\wedge \ov l\psi\bigr)
+mc^2\ov\psi \psi\ov \eta.\label{4.30a}
\end{gather}

We get a new term
\beq{4.29}
-i\,\frac{2\sqrt{G_N}}c\,\hbar F^\m\ov\psi \g_5\si_\m\psi.
\end{equation}
It is an interaction of the \elm c field with an anomalous dipole electric
moment. For such an anomalous interaction it reads
\beq{4.30}
i\,\frac{d_{kk}}2 \,F^\m \ov\psi\g_5\si_\m\psi.
\end{equation}
Our anomalous moment reads
\beq{4.31}
d_{kk}=-\frac{4\sqrt{G_N}}c\,\hbar=-\frac{4l\pl}{\sqrt\a}\,q
\simeq -7.56784835\times 10^{-32}\,{\rm[cm]}q
\end{equation}
where $l\pl$ is a \Pl\ length
$$
l\pl=\sqrt{\frac{\hbar G_N}{c^3}}\simeq1.61199\times10^{-35}{\rm m},
$$
$q$ is an elementary charge and
$$
\a=\frac{e^2}{\hbar c}\simeq\frac1{137}
$$
is a fine structure \ct.

This term can be also rewritten in a different way,
\beq{4.32}
-\frac2{\La_p}\,(\hbar^3c^5)^{1/2}F^\m \ov\psi\g_5\si_\m\psi
\end{equation}
where
\beq{4.33}
\bga
\La_p=m_pc^2\simeq 1.2209\times10^{19}{\rm GeV}\\
m_p=2.1765\times10^{-8}{\rm kg}
\ega
\end{equation}
are \Pl\ energy scale and \Pl\ mass. Thus we get a term which probably gives
us a trace of New Physics on a \Pl\ energy scale. This term is
nonrenormalizable in Quantum Field Theory and it is of 5 order in mass units
(i.e.\ $c=\hbar=1$) divided by an energy (mass) scale.

The term \eqref{4.30} can be written in a very convenient way
\beq{4.34}
d_{kk}\ov\psi\bigl(\b(\os\Si\cdot \os E+i\os\a\os B)\psi\bigr)
\end{equation}
where
\bg{4.35}
\b=\left(\begin{matrix} I & 0 \\ 0 & -I\end{matrix}\right), \quad
\os \a=\left(\begin{matrix} 0 & \os\si \\ \os\si & 0 \end{matrix}\right), \q
\os\g=\b\os\a\\
\os\Si=-\g^5\os \a=\g^4\g^5\os\g=\b\g^5\os \g \label{4.36}\\
\os\si=(\si_x,\si_y,\si_z), \label{4.37}
\end{gather}
$I$ is the identity matrix $2\times2$ and $\os\si$ are Pauli matrices. $\os E$ is an
electric field and $\os B$ is a magnetic field. In this way our term
introduces an anomalous dipole electric interaction and also an anomalous
magnetic dipole interaction. Of course the magnetic interaction is negligible
in comparison to ordinary magnetic moment interaction of an electron. One can
easily calculate this anomalous magnetic moment of an electron in terms of
Bohr magneton getting
$$
\frac4{\sqrt \a}\Bigl(\frac{m_{\rm e}}{m_{\rm p}}\Bigr)\mu_B=19.188
\times10^{-22}\mu_B,
$$
where $m_{\rm e}$ is a mass of an electron and $\mu_B=\frac{q\hbar}{2m_e}$ is
a Bohr magneton. From
the physical point of view the most important is the electric dipole moment
(EDM). So we see that using spinors $\Ps$ and $\ov\Ps$ and a \dv\ $\wt{\ov
\cD}$ in the Kaluza--Klein Theory we have achieved an additional \gr al-\elm
c effect. It is just an existence of a dipole moment of a fermion, which
value is determined by fundamental \ct s (only!).
This is another ``interference effect'' between \elm c and \gr al fields in
our unified field theory.
Thirring also has achieved in his paper \cite{421} a dipole electric moment
of fermion of the same order. In his theory a minimal rest mass of fermion is
of order of a \Pl\ mass. Thus his theory cannot describe a fermion from the
Standard Model. The anomalous moment in Thirring's theory depends on a mass
of a fermion. In order to get $d_{kk}$ of order $10^{-32}\,{\rm[cm]}q$ this must
be of a \Pl\ mass order. Otherwise the value of $d_{kk}$ can be smaller. (In
reality W.~Thirring obtains two types of anomalous Pauli terms---electric and
magnetic of the same order.)

In our case mass $m$ may be arbitrary, e.g.\ $m=0$. Thus we can consider also
massless fermions. We can also consider chargeless fermions, i.e.\ for $k=0$.
It is also worth noticing that Thirring's quantities $\Ps$
and~$\ov\Ps$ have nothing to do with our spinor fields $\Ps$ and~$\ov\Ps$ for
a mysterious Thirring's quantity $\vf$ which is absent in our theory (it
appears also in Thirring's definition of a parity operator). We develop the
theory considered here also in ordinary Kaluza--Klein Theory and in the
Kaluza--Klein theory with a torsion (see Refs \cite{98}, \cite{422},
\cite{423}). Someone develops a theory using our spinors $\Ps$ and~$\ov\Ps$
getting also anomalous electric dipole moments (see Refs \cite{424},
\cite{425}). We develop a similar approach for a Rarita--Schwinger field (see
Ref.~\cite{426}). In the case of the \NK{} we consider also a different
approach (see \cite{427}, \cite{428}). However now we consider the present as
appropriate.

Let us consider operations of reflection defined on a manifold~$P$. To define
them we choose first a local \cd\ system on~$P$ in such a way that we pass
from $\t^A$ to $dx^A$, i.e.\ $(\pi^\ast(dx^\a),dx^5)$. In
this way
\beq{4.38}
x^A=(x^\a,x^5),\quad x^\a=(\os x,t).
\end{equation}
Then
\beq{4.39}
\Ps(p)=\Ps(x^A)=\Ps\bigl((\os x,t),x^5\bigr)
\end{equation}
and we define transformations: space reflection $P$ (do not confuse with a
manifold~$P$), time reversal $T$, charge reflection~$C$ and combined
transformations $PC$, $\t=PCT$,
\beq{4.40}
\Ps^C(x^\a,x^5)=C\Ps^\ast (x^\a,-x^5),
\end{equation}
where $C^{-1}\g_\mu C=-\g_\mu^\ast$.

Taking a section $f$ we get
\beq{4.41}
(\psi^f)^C (x^\a)=C\psi^{f\ast}(x^\a)
\end{equation}
and a charge changes the sign. The \rf\ $x^5\to -x^5$ as a charge \rf\ has
been already suggested by J.~Rayski (see Ref.~\cite{429}). For the space \cd\
\rf\ we have
\beq{4.42}
\Ps^P(x^\a,x^5)=\g^4 \Ps(-\os x,t,x^5).
\end{equation}
Taking a section $f$ we obtain
\beq{4.43}
(\psi^f)^P(\os x,t)=\g^4\psi^f(-\os x,t),
\end{equation}
i.e.\ a normal parity operator on~$E$.

This contrasts with Thirring's definition of the parity operator (Thirring
was forced to change the definition of the parity operator on 5-\di al space
and he could not obtain a normal parity operator on~$E$). The \tf\ of
time-reversal $T$ is defined by
\beq{4.44}
\Ps^T(\os x,t,x^5)=C^{-1}\g^1\g^2\g^3\Ps^\ast(\os x,-t,-x^5).
\end{equation}
Taking a section $f$ we get
\beq{4.45}
(\psi^f)^T(\os x,t)=C^{-1}\g^1\g^2\g^3(\psi^f)^\ast(\os x,-t)
\end{equation}
and a charge does change sign, i.e.\ a normal time-reversal operator on a
spsce-time.

To define a \tf\ $\t=PCT$ we write
\beq{4.46}
\Ps^\t (\os x,t,x^5)=-i\g^5\Ps(-\os x,-t,-x^5).
\end{equation}
Taking a section $f$ we get
\beq{4.47}
(\psi^f)^\t (\os x,t)=-i\g^5\psi^f(-\os x,-t)
\end{equation}
and a charge changes the sign.
The \tf\ $PC$ is as follows
\beq{4.48}
\Ps^{PC} (\os x,t,x^5)=\g^4C\Ps^\ast(-\os x,t,x^5).
\end{equation}
Taking a section $f$ we have
\beq{4.49}
(\psi^f)^{PC} (\os x,t)=\g^4C(\psi^f)^\ast(-\os x,t)
\end{equation}
and a charge changes a sign.

It is clear now that the \tf s obtained by us do not differ from those known
from the literature.

The additional term in Lagrangian \eqref{4.27} breaks $PC$ or $T$ symmetries
as in Thirring's theory (see Ref.~\cite{421}), but Thirring defines the
operator $PC$ in a different way. This can be easily seen by acting on both
sides of Eq.~\eqref{4.29} with the operator defined by Eq.~\eqref{4.48}. Of
course this breaking is very weak and it cannot be linked to $CP$-breaking
term in Cabbibo--Kobayashi--Maskava matrix. From this breaking due to
$\d_{PC}$-phase, which is responsible for $PC$ nonconservation in $K^0,\ov
K{}^0$ mesons decays and also for $D^0,\ov D{}^0$, $B_s,\ov B_s$, $B^0, \ov
B{}^0$ and so on, see Ref.~\cite{430n}, we can get a dipole electric moment of
an electron of order $8\times10^{-41}\,{\rm [cm]}q$ (if there is not New
Physics beyond SM, see Ref.~\cite{430}). This is because all Feynman diagrams
which induce EDM of electron vanish to three loops order.

According to Ref.~\cite{430} electron EDM
\beq{4.46d}
d_e=\biggl(\frac{g^2_w}{32\pi^2}\biggr)
\biggl(\frac{m_e}{M_w}\biggr)
\biggl[\ln\frac{\La^2}{M_W^2}+O(1)\biggr]\,d_W
\end{equation}
where
\beq{4.47d}
d_W=J\biggl(\frac{g_W^2}{32\pi^2}\biggr)\biggl(\frac q{2M_W}\biggr)
\frac{m_b^4m_s^2m_c^2}{M_W^2}
\end{equation}
is EDM for a $W$ boson, $\La$ is an energy scale for a New Physics
(beyond~SM),
$$
J=s_1^2s_2s_3c_1c_2c_3\sin\d_{CP}=2.96\times10^{-5}
$$
(see Ref.~\cite{432n}) is a Jarlskog invariant, $m_b,m_s,m_c$ are masses of quarks (we suppose the
existence of three families of fermions in~SM) and $s_i=\sin\t_i$,
$c_i=\cos\t_i$, $i=1,2,3$.

EDMs of an electron $d_e$ and quarks can induce EDMs of paramagnetic and
diamagnetic atoms
\bea{4.48d}
d_{\rm para}&\sim& 10\a^2Z^3d_e\\
d_{\rm dia}&\sim& 10Z^2\Bigl(\frac{R_N}{R_A}\Bigr)^2 \wt d_q. \label{4.49d}
\end{eqnarray}

For Thalium (Tl) and for Mercury (Hg) one gets
\bea{4.50}
d_{\rm Tl}&=&-585d_e\\
d_{\rm Hg}&=&7\times10^{-3}e(\wt d_u-\wt d_d)+10^{-2}d_e. \label{4.51}
\end{eqnarray}
For a neutron
$$
d_{\rm n}=(1.4\mp0.6)(d_d-0.25d_u)+(1.1\pm0.5)q(\wt d_d+0.5\wt d_u)
$$
where $d_d,d_u$ are EDM of quarks and $\wt d_d,\wt d_u,\wt d_q$ are color EDM
operators (see Ref.~\cite{431} and references cited therein). Recently we
have an upper bound on EDMs (see Ref.~\cite{432} and references cited
therein)
$$
|d_{\rm n}|<2.9\times 10^{-26}\,{\rm[cm]}q, \quad
|d_e|<1.6 \times 10^{-27}\,{\rm [cm]}q, \q
d({}^{199}{\rm Hg})<3.1\times 10^{-29}\,{\rm[cm]}q.
$$
In the case of $\t$-term in QCD we have also $d_n=3\times 10^{-16}
\t\,{\rm [cm]}q$ (see Ref.~\cite{431}).

Recently there has been a significant progress in obtaining an upper limit on
the EDM of an electron by using a polar molecule thorium monoxide (ThO). The
authors of Ref.~\cite{448} obtained an upper limit on~$d_e$,
\beq{4.58}
|d_e|<8.7 \times 10^{-29}{\rm[cm]}q.
\end{equation}
This is only of three orders of magnitude bigger than our result (see
Eq.~\eqref{4.31}). From the other side there is also a progress in
calculation of SM prediction of EDM for an electron coming from a phase
$\d_{CP}$ of CKM matrix. This calculation gives us the so called
\ti{equivalent} EDM (see Ref.~\cite{449}),
\beq{4.59}
d_e^{\rm equiv} \sim 10^{-38}{\rm[cm]}q,
\end{equation}
which is bigger of three orders of magnitude than the result from
Ref.~\cite{430n}.  Moreover, still smaller of six orders than our result. The
parameter~$\t$ from QCD is unknown and has no influence on EDM of an
electron. The existence of EDM of an electron coming from Kaluza--Klein
theory can help us in understanding of an asymmetry of matter-antimatter in the
Universe. This EDM moment which breaks PC and T~symmetry in an explicit way
can have an influence on the surviving of an annihilation matter with
antimatter following Big Bang.

It is interesting to notice that EDM from Kaluza--Klein Theory is the same
for a muon (a $\mu$ meson) and a tauon (a $\tau$ meson)
as for an electron. We get the same value for flavour
states of neutrinos. Due to this, EDM of this value can influence
oscillations of neutrinos species (see Ref.~\cite{450}).

To be honest, we write down a different, however trivial, coupling of spinor
fields $\Ps$ and~$\ov\Ps$ in Kaluza--Klein. This is a coupling to a \cn\ of
the form
\beq{4.60}
\gd\wh w{},A,B,=\left(
\begin{array}{c|c}
\pi^\ast\bigl(\gd{\wt{\ov w}}{},\a,\b,)\, &\,0\\[2pt]\hline
0 &\, 0
\end{array}
\right).
\end{equation}
In this way $\Ps$ and $\ov\Ps$ are transforming according to ${\rm SL}(2,\C)$
and new phenomena are absent, i.e.\ we have to do with Lagrangian
\eqref{4.28}.

Let us come back to neutrino oscillations in the presence of EDM. Let us
write a Lagrangian for three neutrino species neglecting \gr al field:
\bml{4.61}
\cL_D(\Ps_\l,\ov\Ps_\l,d)=\sum_{\l=\a,\b,\g}\biggl(\frac{i\hbar c}2
\bigl(\ov\Ps_\l l\wedge d\Ps_\l+ d\ov\Ps_\l\wedge l\Ps_\l\bigr)
+i\,\frac{d_{kk}}2 \,F^\m \ov\Ps_\l \g_5\si_\m \Ps_\l\biggr)\\
{}+\sum_{\l,\l'=\a,\b,\g}c^2\ov\Ps_\l m_{\l\l'}\Ps_{\l'}\eta.
\end{multline}
Despite the smallness of $d_{kk}$ its interaction with a strong electric and
magnetic fields can result in sizeable effects (see Eq.~\eqref{4.34}).
$m_{\l\l'}$~is a mass matrix for neutrinos which is not diagonal. In
particular $\a=e$, $\b=\mu$, $\g=\tau$.

Let us consider mass eigenstates of our neutrinos $\Ps_a$, $a=1,2,3$ (see
\cite{450})
\beq{4.62}
\PS_\l=\sum_{a=1,2,3}U_{\l a}\Ps_a.
\end{equation}
The unitary matrix $U=(U_{\l a})$ diagonalizes the mass matrix $\ov
m=(m_{\l\l'})$. The eigenvalues of the mass matrix are called $m_a$, $a=1,2,3$.
\begin{gather}\label{4.63}
\left(\begin{matrix}
m_1 & 0 & 0\\ 0 &m_2& 0\\ 0 &0 &m_3
\end{matrix}\right)=U^+ \,\ov m\,U\\
U=\left(\begin{matrix}
c_{12}c_{13} & s_{12}c_{13} & s_{13}e^{-i\d}\\
-s_{12}c_{23}-c_{12}s_{23}s_{13}e^{i\d} &
c_{12}c_{23}-s_{12}s_{23}s_{13}e^{i\d} & s_{23}c_{13}\\
s_{12}s_{23}-c_{12}c_{23}s_{13}e^{i\d} &
-c_{12}s_{23}-s_{12}c_{23}s_{13}e^{i\d} & c_{23}c_{13}
\end{matrix}\right)
\label{4.64}
\end{gather}
where $c_{ij}=\cos\t_{ij}$, $s_{ij}=\sin\t_{ij}$, the angles $\t_{ij}
\in[0,\frac\pi2]$, $\d\in[0,2\pi]$ is a Dirac CP violation phase (see
Refs.~\cite{430n}, \cite{450}, $i$ means $\l$---flavour, $j$ means $a$---mass
eigenstate).

In the new spinor variables the Lagrangian \eqref{4.61} reads
\beq{4.65}
\cL_D(\Ps_a,\ov\Ps_a,d)=\sum_{a=1,2,3}\biggl(\frac{i\hbar c}2 \bigl(\ov\Ps_a l\wedge d\Ps_a
+d\ov\Ps_a\wedge l\Ps_a\bigr) + \ov\Ps_a M_a \Ps \eta\biggr),
\end{equation}
where
\beq{4.66}
M_a=m_ac^2 + i\,\frac{d_{kk}}2 F^\m \g_5\si_\m = m_ac^2+d_{kk}\b
\bigl(\os\Si\cdot \os E+ i\os\a \cdot\os B\bigr)
\end{equation}
(see Eq.~\eqref{4.34}).

Using initial conditions for mass eigenstates
\bg{4.67}
\Ps_a(\os r,t=0)=\Ps_a^{(0)}(\os r) \\
\Ps_\la^{(0)}(\os r) = U_{\la a}\Ps_a^{(0)}(\os r)\label{4.68} \\
\Ps_a^{(0)}(\os r) =(U^{-1})_{a\la}\Ps_\la^{(0)}(\os r) \label{4.69}
\end{gather}
we can solve an initial value problem for linear \e s corresponding to the
Lagrangian \eqref{4.65}, finding an evolution in time of fields $\Ps_a$ (they
do not couple). Afterwards using \eqref{4.62} and \eqref{4.69} we find
oscillations of three neutrino flavours under an influence of magnetic and
electric fields due to additional term coming from Kaluza--Klein Theory.
Field \e s for $\Ps_a$ (Euler--Lagrange \e s for Lagrangian \eqref{4.65}) are
given in the following Hamilton form
\beq{4.70}
i\hbar c\,\frac{\pa \PS_a}{\pa t}=H_a\Ps_a,\quad a=1,2,3,
\end{equation}
where
\bg{4.71}
H_a=c\os\a\cdot\os p+\b m_ac^2-d_{kk}\bigl(\os\Si\cdot\os E+i\os\a\cdot
\os B\bigr)\\
\os p=-i\hbar\os\nabla.\label{4.72}
\end{gather}
Thus eventually one gets
\beq{4.73}
i\hbar c\,\frac{\pa \Ps_a}{\pa t}=-i\hbar c(\os\a\cdot\os\nabla)\Ps_a
+m_ac^2 \b\Ps_a- d_{kk}(\os\Si\cdot\os E+i\os \a\cdot\os B)\Ps_a, \q
a=1,2,3.
\end{equation}

Equations \eqref{4.73} are typical Dirac--Pauli \e s. Moreover, they have a
term which explicitly breaks PC \tf. We suppose $\os E={\rm const}$, $\os
B={\rm const}$. For Eqs~\eqref{4.73} are linear the general \so s are
expressed by the Fourier integral
\bml{4.74}
\Ps_a(\os r,t)=\int \frac{d^3\os p}{(2\pi)^{3/2}} \,e^{i\os p\cdot\os r}\\
{}\times \sum_{\z=\pm1}\Bigl[a_a^{(\z)} u_a^{(\z)}(\os p)
\exp\bigl(-iE(+)_a^{(\z)}t\bigr) + b_a^{(\z)}v_a^{(\z)}(\os p)
\exp\bigl(-i E(-)_a^{(\z)}t\bigr)\Bigr]
\end{multline}
where $a_a^{(\z)},b_a^{(\z)}$ are arbitrary \cf s, $u_a^{(\z)},v_a^{(\z)}$
are base spinors \st
\bea{4.75}
H_a u_a^{(\z)}&=&E(+)_a^{(\z)}u_a^{(\z)}\\
H_a v_a^{(\z)}&=&E(-)_a^{(\z)}v_a^{(\z)}.\label{4.76}
\end{eqnarray}
In the classical situation
\beq{4.77}
E(+)_a^{(\z)}=-E(-)_a^{(\z)}
\end{equation}
and $\z=\pm1$ describes different polarization states of the fermions $\Ps_a$
(see Refs \cite{1*}, \cite{**}). In our case $E(+)_a^{(+1)}$, $E(+)_a^{(-1)}$,
$E(-)_a^{(-1)}$, $E(-)_a^{(+1)}$ are roots of the polynomial of the fourth
order
\beq{4.78}
\det(H_a(\os p)-IE_a)=0, \q a=1,2,3,
\end{equation}
where $I$ is the identity matrix $4\times4$ and
\beq{4.79}
H_a(\os p)=c\os \a\cdot\os p+\b m_ac^2-d_{kk}\bigl(\os\Si\cdot\os E
+i\os\a\cdot\os\b\bigr), \q a=1,2,3.
\end{equation}

Spinors $u_a^{(\z)},v_a^{(\z)}$ are eigenvectors corresponding to those
eigenvalues. They are orthogonal.
Using formulae \eqref{4.35}--\eqref{4.37} one transforms Eqs
\eqref{4.78}--\eqref{4.79} into
\beq{4.80a}
H_a=\left(\begin{matrix}
m_ac^2I - d_{kk}(\os E\cdot\os \si) &&(c\os p-id_{kk}\os B)\cdot\os\si\\
(c\os p-id_{kk}\os B)\os\si &&d_{kk}(\os E\cdot\os \si)-m_ac^2I
\end{matrix}\right), \q a=1,2,3,
\end{equation}
and
\beq{4.81a}
\det\left(\begin{matrix}
(m_ac^2-E_a)I - d_{kk}(\os E\cdot\os \si) &&(c\os p-id_{kk}\os B)\cdot\os\si\\
(c\os p-id_{kk}\os B)\os\si &&d_{kk}(\os E\cdot\os \si)-(m_ac^2+E_a)I
\end{matrix}\right)=0, \q a=1,2,3,
\end{equation}
where $I$ is the $2\times2$ identity matrix.

Using explicit forms of Pauli matrices
\beq{4.82a}
\si_x=\left(\begin{matrix} 0 & 1 \\ 1 & 0 \end{matrix}\right)\q
\si_y=\left(\begin{matrix} 0 & -i \\ i & 0 \end{matrix}\right)\q
\si_z=\left(\begin{matrix} 1 & 0 \\ 0 & -1 \end{matrix}\right)
\end{equation}
one eventually gets
\beq{4.83a}
\hskip-6pt\det\left(\begin{matrix}
\gathered m_ac^2-E_a\\{}-d_{kk}E_z\endgathered &&
-d_{kk}(E_x-iE_y)&&
cp_z-id_{kk}B_z&&
\gathered c(p_x-ip_y)\\{}-id_{kk}(B_x-iB_y)\endgathered \\
\noalign{\vskip8pt}
-d_{kk}(E_x+iE_y)&&
\gathered m_ac^2-E_a\\{}+d_{kk}E_z\endgathered &&
\gathered c(p_x+ip_y)\\{}-id_{kk}(B_x+iB_y)\endgathered&&
-cp_z+id_{kk}B_z\\
\noalign{\vskip8pt}
cp_z-id_{kk}B_z&&
\gathered c(p_x-ip_y)\\{}-id_{kk}(B_x-iB_y)\endgathered&&
\gathered -m_ac^2-E_a\\{}+d_{kk}E_z\endgathered &&
d_{kk}(E_x-iE_y)\\
\noalign{\vskip8pt}
\gathered c(p_x+ip_y)\\{}-id_{kk}(B_x+iB_y)\endgathered&&
-cp_z+id_{kk}B_z&&
d_{kk}(E_x+iE_y)&&
\gathered -m_ac^2-E_a\\{}+d_{kk}E_z\endgathered
\end{matrix}\right)=0.
\end{equation}

Using initial conditions we can determine
\cf s $a_a^{(\z)}$ and~$b_a^{(\z)}$, i.e.\ we expand $\Ps_a^{(0)}(\os r)$
into Fourier integral
\beq{4.80}
\Ps_a^{(0)}(\os r)=\int \frac{d^3\os p}{(2\pi)^{3/2}}\,e^{i\os p\cdot\os r}
\sum_{\z=\pm1}\bigl[a_a^{(\z)}u_a^{(\z)}(\os p)+b_a^{(\z)}v_a^{(\z)}(\os p)
\bigr], \q a=1,2,3.
\end{equation}
We can consider several possibilities of neutrino flavour oscillations
supposing e.g.
\beq{4.81}
\Ps_\a^{(0)}(\os r)=\xi(\os r) \qh{and}
\Ps_\b^{(0)}(\os r)=\Ps_\g^{(0)}(\os r)=0.
\end{equation}
In this way
\beq{4.82}
\Ps_a^{(0)}(\os r)=U_{a\a}\xi(\os r)
\end{equation}
which can be considered as initial conditions for oscillations.

Moreover, this problem is beyond the scope of this paper and will be
considered elsewhere.

Let us notice that our generalization of a minimal coupling scheme
Eq.~\eqref{4.22} induces a new \cn\ on~$P$.
\bg{4.83}
\gd\check w{},A,B,=\hor(\gd\wt w,A,B,)\\
\qh{or} \gd\check w{},A,B,=\left(
\begin{array}{c|c}
\pi^\ast\bigl(\gd{\wt{\ov w}}{},\a,\b,)\, &\,
\frac\l2 \pi^\ast(\gd F,\a,\g,\ov \t{}^\g)\\[2pt]\hline
\noalign{\vskip2pt}
-\frac\l2\pi^\ast (F_{\b\g}\ov \t{}^\g) & 0
\end{array}
\right).\label{4.84}
\end{gather}
This \cn\ is metric but with non-vanishing torsion. Properties of this \cn\
have been extensively examined (also in the case of nonabelian Kaluza--Klein
Theories) in Ref.~\cite{0}.

Let us consider the following problem. What would it mean for Physics if
someone measured an  EDM for an electron of the value $d_{kk}=-\frac{4l_{\rm
pl}}{\sqrt\a}q$ as predicted in this paper?
It would mean the fifth dimension is a reality in the sense of a 5-dimensional
Minkowski space.

An experiment which measures such a quantity strongly supports an idea of
rotations around the fifth axis in this space (the fifth dimension is a
space-like). This EDM exists only due to these rotations. Otherwise spinor
fields couple to a \cn\ \eqref{4.60} and there is not a new effect.

Even $P$ is a 5-\di al manifold, the additional fifth \di\ is not necessarily
of the same nature as the remaining four \di s, in particular three space \di
s. This \di\ is a gauge \di\ connected to the \elm c field. Moreover, we can
develop this theory using Yang--Mills' fields and also Higgs' fields using
\di al reduction procedure, expecting some additional effects. It means we
can expect something as ``travelling'' along additional \di s. This
perspective would have a tremendous importance for Physics and Technology.

Simultaneously an existence of an EDM of an electron has also very great
impact on our understanding of PC and~T symmetries breaking. This is also
very important.

Thus a mentioned measurement with an answer: \ti{Yes}, would have very
important physical, technological and even philosophical implications.

Some recent proposals for measuring of EDM for an electron can be found
in Refs \cite{a,b,c}. In Ref.~\cite{a} there is a proposal with the upper
limit of $5\times10^{-30}\ {\rm e\cdot cm}$.

The anomalous interaction term \eqref{4.29} can be treated together with gravity
(\nos) and \elm sm using nonlocal quantization procedure similar to ideas from
Ref.~\cite{q} using methods from Refs \cite{al,be,ga,de,p,w}. It means we use
this term as a source term in the \NK{} (see Ref.~\cite{q} and the references
therein).

\section*{Appendix A}
\def\theequation{A.\arabic{equation}}
\setcounter{equation}0
In the appendix we describe the notation and definitions of geometric
quantities used in the paper. We use a smooth principal bundle which is an
ordered sequence
\beq{A.0}
\ul P=(P,F,G,E,\pi),
\end{equation}
where $P$ is a total bundle manifold, $F$ is typical fibre, $G$, a Lie group,
is a structural group, $E$~is a base manifold and $\pi$ is a projection. In our
case $G=\U(1)$, $E$~is a \spt, $\pi:P\to E$.
We have a map $\vf:P\times G\to P$ defining an
action of~$G$ on~$P$. Let $a,b\in G$ and $\ve$~be a unit element of the
group~$G$, then $\vf(a)\circ \vf(b)=\vf(ba)$, $\vf(\ve)=\id$, where $\vf(a)p
=\vf(p,a)$. Moreover, $\pi\circ\vf(a)=\pi$. For any open set $U\subset E$ we
have a local trivialization $U\times G\simeq \pi^{-1}(U)$. For any $x\in E$,
$\pi^{-1}(\{x\})=F_x \simeq G$, $F_x$ is a fibre over~$x$ and is equal to~$F$.
In our case we suppose $G=F$, i.e.\ a Lie group $G$ is a typical fibre.
$\o$~is a 1-form
of \cn\ on~$P$ with values in the algebra of~$G$, $\mathfrak G$. In the case
of $G=\U(1)$ we use a notation $\a$ (an \elm c \cn). Lie algebra of $U(1)$
is~$R$. Let $\vf'(a)$ be a
tangent map to $\vf(a)$ whereas $\vf^\ast(a)$ is the contragradient
to~$\vf'(a)$ at a point~$a$. The form $\o$ is a form of ad-type, i.e.
\beq{A.1}
\vf^\ast(a)\o=\ad{a^{-1}}'\o,
\end{equation}
where $\ad{a^{-1}}'$ is a tangent map to the internal automorphism of the
group~$G$
\beq{A.2}
\ad a(b)=aba^{-1}.
\end{equation}
In the case of $\U(1)$ (abelian) the condition \eqref{A.1} means
\beq{A.3}
\mathop{\cL}_{\z_5}\a=0,
\end{equation}
where $\z_5$ is a Killing vector corresponding to one generator of the group
$\U(1)$. Thus this is a vector tangent to the operation of the group $\U(1)$
on~$P$, i.e.\ to $\vf_{\exp(i\chi)}$, $\chi=\chi(x)$, $x\in E$,
$\mathop{\cL}\limits_{\z_5}$ is a Lie \dv\ along $\z_5$.
We may introduce the distribution (field) of linear elements $H_r$, $r\in P$,
where $H_r\subset T_r(P)$ is a subspace of the space tangent to~$P$ at a
point~$r$ and
\beq{A.4}
v\in H_r \iff \o_r(v)=0.
\end{equation}
So
\beq{A.5}
T_r(P)=V_r\oplus H_r,
\end{equation}
where $H_r$ is called a subspace of \ti{horizontal\/} vectors and $V_r$ of
\ti{vertical\/} vectors.
For vertical vectors $v\in V_r$ we have $\pi'(v)=0$. This means that $v$ is
tangent to the fibres.

Let
\beq{A.6}
v=\hor(v)+\ver(v),\quad \hor(v)\in H,\ \ver(v)\in V_r.
\end{equation}
It is proved that the distribution $H_r$ is equal to choosing a \cn~$\o$. We
use the operation $\hor$ for forms, i.e.
\beq{A.7}
(\hor\b)(X,Y)=\b(\hor X,\hor Y),
\end{equation}
where $X,Y\in T(P)$.

The 2-form of a curvature is defined as follows
\beq{A.8}
\O=\hor d\o=D\o,
\end{equation}
where $D$ means an exterior covariant \dv\ \wrt $\o$. This form is also of
ad-type.

For $\O$ the structural Cartan \e\ is valid
\beq{A.9}
\O=d\o+\tfrac12[\o,\o],
\end{equation}
where
\beq{A.10}
[\o,\o](X,Y)=[\o(X),\o(Y)].
\end{equation}
Bianchi's identity for $\o$ is as follows
\beq{A.11}
D\O=\hor d\O=0.
\end{equation}
The map $f:E\supset U\to P$ such that $f\circ \pi=\id$ is called a
\ti{section} ($U$ is an open set).

From physical point of view it means choosing a gauge. A~covariant \dv\
on~$P$ is defined as follows
\beq{A.12}
D\Ps=\hor d\Ps.
\end{equation}
This \dv\ is called a \ti{gauge \dv}. $\Ps$ can be a spinor field on~$P$.

In this paper we use also a linear \cn\ on manifolds $E$ and $P$, using the
formalism of differential forms. So the basic quantity is a one-form of the
\cn\ $\gd\o,A,B,$. The 2-form of curvature is as follows
\beq{A.13}
\gd\O,A,B,=d\gd\o,A,B,+\gd\o,A,C, \wedge \gd\o,C,B,
\end{equation}
and the two-form of torsion is
\beq{A.14}
\T^A=D\t^A,
\end{equation}
where $\t^A$ are basic forms and $D$ means exterior covariant \dv\ \wrt \cn\
$\gd\o,A,B,$. The following relations are established \cn s with generally met
symbols
\beq{A.15}
\bal
\gd\o,A,B,&=\gd\G,A,BC,\t^C\\
\T^A&=\tfrac12\gd Q,A,BC,\t^B\wedge \t^C\\
\gd Q,A,BC,&=\gd\G,A,BC,-\gd\G,A,CB,\\
\gd\O,A,B,&=\tfrac12 \gd R,A,BCD,\t^C \wedge \t^D,
\eal
\end{equation}
where $\gd\G,A,BC,$ are \cf s of \cn\ (they do not have to be \s\ in indices
$B$ and~$C$), $\gd R,A,BCD,$ is a tensor of a curvature, $\gd Q,A,BC,$ is a
tensor of a torsion in a holonomic frame. Covariant exterior derivation \wrt $\gd\o,A,B,$ is given
by the formula
\beq{A.16}
\bal
D\Xi^A&=d\Xi^A+\gd\o,A,C,\wedge \Xi^C\\
D\gd\Si,A,B,&=
d\gd\Si,A,B,+\gd\o,A,C,\wedge \gd\Si,C,B,-\gd\o,C,B,\wedge \gd\Si,A,C,.
\eal
\end{equation}
The forms of a curvature $\gd\O,A,B,$ and torsion $\T^A$ obey Bianchi's
identities
\beq{A.17}
\bal
{}&D\gd\O,A,B,=0\\
&D\T^A=\gd\O,A,B,\wedge \t^B.
\eal
\end{equation}
All quantities introduced here can be found in Ref.~\cite{KN}.

In this paper we use a formalism of a fibre bundle over a \spt~$E$ with an
\elm c \cn~$\a$ and traditional formalism of differential geometry for linear
\cn s on~$E$ and~$P$. In order to simplify the notation we do not use fibre
bundle formalism of frames over $E$ and~$P$. A~vocabulary connected geometrical
quantities and gauge fields (Yang--Mills fields) can be found in
Ref.~\cite{97}.

In Ref.~\cite{Wu} we have also a similar vocabulary (see Table~I, Translation
of terminology). Moreover, we consider a little different terminology. First
of all we distinguished between a gauge \pt\ and a \cn\ on a fibre bundle. In
our terminology a gauge \pt\ $A_\mu \ov\t{}^\mu$ is in a particular gauge $e$
(a~section of a bundle), i.e.
\beq{A.18}
A_\mu \ov\t{}^\mu=e^\ast\o
\end{equation}
where $A_\mu \ov\t{}^\mu$ is a 1-form defined on $E$ with values in a Lie
algebra $\mathfrak G$ of~$G$. In the case of a strength of a gauge field we have
similarly
\beq{A.19}
\tfrac12 F_\m \ov\t{}^\mu \wedge \ov\t{}^\nu=e^\ast\O
\end{equation}
where $F_\m \ov\t{}^\mu \wedge \ov\t{}^\nu$ is a 2-form defined on~$E$ with
values in a Lie algebra $\mathfrak G$ of~$G$.

Using generators of a Lie algebra $\mathfrak G$ of $G$ we get
\beq{A.20}
A=\gd A,a,\mu, \ov\t{}^\mu X_a=e^\ast \o \quad\hbox{and}\quad
F=\tfrac12\gd F,a,\m,\ov\t{}^\mu \wedge \ov\t{}^\nu X_a=e^\ast \O
\end{equation}
where
\beq{A.21}
[X_a,X_b]=\gd C,c,ab,X_c, \quad a,b,c=1,2,\dots,n, \ n=\dim G(=\dim \mathfrak G),
\end{equation}
are generators of $\mathfrak G$, $\gd C,c,ab,$ are structure \ct s of a Lie
algebra of~$G$, $\mathfrak G$, $[\cdot,\cdot]$ is a commutator of Lie algebra
elements.

In this paper we are using Latin lower case letters for 3-\di al space indices. Here
we are using Latin lower case letters as Lie algebra indices. It does not
result in any misunderstanding.
\beq{A.22}
\gd F,a,\m,=\pa_\mu\gd A,a,\nu,-\pa_\nu\gd A,a,\mu,+\gd C,a,bc,\gd A,b,\mu,
\gd A,c,\nu,.
\end{equation}
In the case of an \elm c \cn\ $\a$ the field strength~$F$ does not depend on
gauge (i.e.\ on a section of a~bundle).

Finally it is convenient to connect our approach using gauge \pt s $\gd
A,a,\mu,$ with usually met (see Ref.~\cite{Pok}) matrix valued gauge
quantities $A_\mu$ and $F_\m$. It is easy to see how to do it if we consider
Lie algebra generators $X_a$ as matrices. Usually one supposes that $X_a$ are
matrices of an adjoint representation of a Lie algebra~$\mathfrak G$, $T^a$
with a normalization condition
\beq{A.23}
{\rm Tr}(\{T^a,T^b\})=2\d^{ab},
\end{equation}
where $\{\cdot,\cdot\}$ means anticommutator in an adjoint representation.

In this way
\bea{A.24}
A_\mu&=&\gd A,a,\mu, T^a,\\
F_\m&=&\gd F,a,\m, T^a. \label{A.25}
\end{eqnarray}
One can easily see that if we take
\beq{A.26}
F_\m=\pa_\mu A_\nu - \pa_\nu A_\mu + [A_\mu,A_\nu]
\end{equation}
from Ref.~\cite{Pok} we get
\beq{A.26a}
F_\m=(\gd F,a,\m,)T^a,
\end{equation}
where $\gd F,a,\m,$ is given by \eqref{A.22}. From the other side if we take
a section $f$, $f:U\to P$, $U\subset E$, and corresponding to it
\bea{A.27}
\ov A=\gd\ov A,a,\mu, \ov\t{}^\mu X_a&=&f^\ast \o\\
\ov F=\tfrac12\gd\ov F,a,\m, \ov\t{}^\mu \wedge \ov\t{}^\nu X_a&=&f^\ast \O
\label{A.28}
\end{eqnarray}
and consider both sections $e$ and $f$ we get transformation from $\gd
A,a,\mu,$ to $\gd\ov A{},a,\mu,$ and from $\gd F,a,\m,$ to $\gd\ov F{},a,\m,$ in
the following way. For every $x\in U\subset E$ there is an element $g(x)\in
G$ such that
\beq{A.29}
f(x)=e(x)g(x)=\vf(e(x),g(x)).
\end{equation}
Due to \eqref{A.1} one gets
\bea{A.30}
\ov A(x)&=&\ad{g^{-1}(x)}'A(x)+{g^{-1}(x)}\,dg(x)\\
\ov F(x)&=&\ad{g^{-1}(x)}'F(x) \label{A.31}
\end{eqnarray}
where $\ov A(x),\ov F(x)$ are defined by \eqref{A.27}--\eqref{A.28} and
$A(x),F(x)$ by \eqref{A.20}. The formulae \eqref{A.30}--\eqref{A.31} give a
geometrical meaning of a gauge transformation (see Ref.~\cite{97}). In an
\elm c case $G=\U(1)$ we have similarly, if we change a local
section from $e$ to~$f$ we get
$$
f(x)=\vf(e(x), \exp(i\chi(x)))  \quad (f:U\supset E\to P)
$$
and $\ov A=A+d\chi$.

Moreover,
in the traditional approach (see Ref.~\cite{Pok}) one gets
\bea{A.32}
\ov A_\mu(x)&=&U(x)^{-1}A_\mu(x)U(x)+U^{-1}(x)\pa_\mu U(x)\\
\ov F_\m(x)&=&U^{-1}(x)F_\m U(x), \label{A.33}
\end{eqnarray}
where $U(x)$ is the matrix of an adjoint representation of a Lie group $G$.

For an action of a group $G$ on $P$ is via \eqref{A.1}, $g(x)$ is exactly a
matrix of an adjoint representation of~$G$. In this way
\eqref{A.30}--\eqref{A.31} and \eqref{A.32}--\eqref{A.33} are equivalent.

Let us notice that usually a Lagrangian of a gauge field (Yang--Mills field)
is written as
\beq{A.34}
\cL_{\rm YM} \sim {\rm Tr}(F_\m F^\m)
\end{equation}
where $F_\m$ is given by \eqref{A.25}--\eqref{A.26}. It is easy to see that
one gets
\beq{A.35}
\cL_{\rm YM} \sim h_{ab}\gd F,a,\m, F^{b\m}
\end{equation}
where
\beq{A.36}
h_{ab}=\gd C,d,ac, \gd C,c,bd,
\end{equation}
is a Cartan--Killing tensor for a Lie algebra $\mathfrak G$, if we remember
that $X_a$ in adjoint representation are given by structure \ct s $\gd
C,c,ab,$.

Moreover, in Refs \cite{1,3} we use the notation
\beq{A.39}
\O=\tfrac12 \gd H,a,\m,\t^\mu \wedge \t^\nu X_a.
\end{equation}
In this language
\beq{A.40}
\cL_{\rm YM}=\tfrac1{8\pi} h_{ab}\gd H,a,\m, H^{b\m}.
\end{equation}
It is easy to see that
\beq{A.41}
e^\ast (\gd H,a,\m,\t^\mu \wedge \t^\nu X_a)=\gd F,a,\m,{\ov\t}^\mu
\wedge {\ov \t}^\nu X_a.
\end{equation}
Thus \eqref{A.40} is equivalent to \eqref{A.35} and to \eqref{A.34}.
\eqref{A.34} is invariant to a change of a gauge. \eqref{A.40} is invariant
\wrt the action of a group~$G$ on~$P$.

Let us notice that $h_{ab}\gd F,a,\m, F^{b\m}=h_{ab}\gd H,a,\m, \gd H,b,\m,$,
even $\gd H,a,\m,$ is defined on~$P$ and $\gd F,a,\m,$ on~$E$. In the
non-abelian case it is more natural to use $\gd H,a,\m,$ in place of $\gd
F,a,\m,$.

\section*{Appendix B}
\def\theequation{B.\arabic{equation}}
\setcounter{equation}0
\def\rp{representation}
\def\SL{{\rm SL}}

In this paper we consider two kinds of spinor fields $\Ps,\ov\Ps$ and
$\psi,\ov\psi$ defined respectively on~$P$ and~$E$. Spinor fields $\Ps$
and~$\ov\Ps$ transform according to $\Spin(1,4)$ and $\psi,\ov\psi$ according
to $\Spin(1,3)\simeq{\rm SL}(2,\C)$. We have
\beq{C.1}
U(g)\Psi(X)=D^F(g)\Psi(g^{-1}X), \q X\in M^{(1,4)}, \ g\in\SO(1,4).
\end{equation}
$\SO(1,4)$ acts linearly in $M^{(1,4)}$ (5-\di al Minkowski space). The
Lorentz group $\SO(1,3)\subset\SO(1,4)$. $D^F$ is a representation of
$\SO(1,4)$ (de~Sitter group) \st after a restriction to its subgroup
$\SO(1,3)$ we get
\beq{C.2}
D^F{}_{|\SO(1,3)}(\La)=L(\La),
\end{equation}
where
\beq{C.3}
L(\La)=D^{(1/2,0)}(\La)\oplus D^{(0,1/2)}(\La)
\end{equation}
is a Dirac representation of $\SO(1,3)$. More precisely, we deal with \rp s
of $\Spin(1,4)$ and $\Spin(1,3)\simeq \SL(2,\C)$ (see Ref.~\cite{52}). In
other words, we want spinor fields $\Ps$ and~$\ov\Ps$ to transform according
to such a \rp\ of $\Spin(1,4)$ which is induced by a Dirac \rp\
of~$\SL(2,\C)$. The complex \di s of both \rp s are the same:~4. The same are
also Clifford algebras
\beq{C.4}
C(1,4)\simeq C(1,3)
\end{equation}
(see Refs \cite{53}, \cite{54}).

One gets (up to a phase)
\beq{C.5}
\Ps_{|\SL(2,\C)}=\psi.
\end{equation}
Spinor fields $\psi$ and $\ov\psi$ transform according to Dirac \rp,
$\ov\psi=\psi^+B$. Our matrices $\g_\mu$ and $\g_A$ are \rp s of $C(1,3)$
($C(1,4)$).  One can consider projective \rp s for $\Ps$ and~$\psi$, i.e.\
\rp s of $\Spin(1,3)\ot \U(1)$ and $\SL(2,\C)\ot \U(1)$. Moreover, we do not
develop this idea here.

In this paper we develop the following approach to spinor fields on~$E$ and
on~$P$. We introduce orthonormal frames on~$E$ ($dx^1$, $dx^2$, $dx^3$,
$dx^4$) and on~$P$ ($dX^1=\pi^\ast(dx^1)$, $dX^2=\pi^\ast(dx^2)$,
$dX^3=\pi^\ast(dx^3)$, $dX^4=\pi^\ast(dx^4)$, $dX^5$). Our spinors $\Ps$
on $(P,\g_\(AB))$ and $\psi$ on $(E,g_\(\a\b))$ are defined as complex bundles
$\C^4$ over~$P$ or~$E$ with homomorphisms $\rho:C(1,4)\to \cL(\C^4)$ (resp.\
$\rho:C(1,3)\to \cL(\C^4)$) of bundles of algebras over~$P$ (resp.~$E$) \st
for every $p\in P$ (resp.~$x\in E$), the restriction of~$\rho$ to the fiber
over~$p$ (resp.~$x$) is equivalent to spinor \rp\ of a Clifford algebra
$C(1,4)$ (resp.~$C(1,3)$), i.e.\ $D^F$ (resp.\ Dirac \rp, see Refs
\cite{55},~\cite{56}). (There is also a paper on a similar subject (see
Ref.~\cite{57}).) Spinor fields $\Ps$ and~$\psi$ are sections of these
bundles. There is also an approach to consider spinor bundles for~$\Ps$
and~$\psi$ as bundles associated to principal bundles of orthonormal frames
for $(P,\g_\(AB))$ or $(E,g_\(\a\b))$ (spin frames). Spinor fields $\Ps$ and
$\psi$ are sections of these bundles. In our case we consider spinor fields
$\Ps$ and~$\ov\PS$ transforming according to \eqref{4.13} and \eqref{4.14}.
In the case of $\psi$ and $\ov\psi$ we have
\beq{C.6}
\bga
\ov\t{}^{\a\prime}=\ov\t{}^\a+\d\ov\t{}^\a=\ov\t{}^\a-\gd\ve,\a,\b,
\ov\t{}^\b\\
\ov\ve_{\a\b}+\ov\ve_{\b\a}=0.
\ega
\end{equation}

If the spinor field $\psi$ corresponds to $\ov\t{}^\a$ and $\psi'$ to
$\ov\t{}^{\a\prime}$ we get
\beq{C.7}
\bal
\psi'&=\psi+\d\psi=\psi-\ov\ve{}^{\a\b}\si_{\a\b}\psi\\
\ov\psi{}'&=\ov\psi+\d\ov\psi=\ov\psi+\ov\psi\ov\ve{}^{\a\b}\si_{\a\b}.
\eal
\end{equation}
Spinor fields $\Ps$ and $\ov\Ps$ are $\psi$ and $\ov\psi$ in any section of a
bundle~$P$. Simultaneously we suppose conditions \eqref{4.2}.

Similarly as for $\Ps,\ov\PS$ one gets
\bg{C.8}
\bal
\wt{\ov D}\psi&=d\psi + \gd\wt{\ov w}{},\a,\b, \dg\si,\a,\b, \psi\\
\wt{\ov D}\,\ov\psi&=d\ov\psi - \gd\wt{\ov w}{},\a,\b, \ov\psi \dg\si,\a,\b,
\eal\\
\bal
\wt{\ov\cD}\psi=\hor\wt{\ov D}\psi=\gdv\psi
+ \gd\wt{\ov w}{},\a,\b,\dg\si,\a,\b,\psi\\
\wt{\ov\cD}\,\ov\psi=\hor\wt{\ov D}\,\ov\psi=\gdv\ov\psi
- \gd\wt{\ov w}{},\a,\b,\ov\psi\dg\si,\a,\b,.
\eal \label{C.9}
\end{gather}

\goodbreak

\end{document}